# BSAGIoT: A Bayesian Security Aspect Graph for Internet of Things (IoT)

Zeinab Lashkaripour[a], Masoud Khosravi-Farmad[b], AhmadReza Montazerolghaem[c*], Razieh Rezaee[d]

[a] Data Science Department, Montreal College of Information Technology, Montreal, Canada. zeinab.lashkaripour@faculty.montrealcollege.ca
[b] Data and Communication Security Lab., Computer Engineering Department, Ferdowsi University of Mashhad, Mashhad, Iran. m.khosravi@mail.um.ac.ir
[c] Faculty of Computer Engineering, University of Isfahan, Isfahan, Iran. a.montazerolghaem@comp.ui.ac.ir
[d] Computer Engineering Department, Imam Reza University, Mashhad, Iran. r.rezaee@imamreza.ac.ir
[*] Corresponding Author

**Abstract:** IoT is a dynamic network of interconnected things that communicate and exchange data, where security is a significant issue. Previous studies have mainly focused on attack classifications and open issues rather than presenting a comprehensive overview on the existing threats and vulnerabilities. This knowledge helps analyzing the network in the early stages even before any attack takes place. In this paper, the researchers have proposed different security aspects and a novel Bayesian Security Aspects Dependency Graph for IoT (BSAGIoT) to illustrate their relations. The proposed BSAGIoT is a generic model applicable to any IoT network and contains aspects from five categories named data, access control, standard, network, and loss. This proposed Bayesian Security Aspect Graph (BSAG) presents an overview of the security aspects in any given IoT network. The purpose of BSAGIoT is to assist security experts in analyzing how a successful compromise and/or a failed breach could impact the overall security and privacy of the respective IoT network. In addition, root cause identification of security challenges, how they affect one another, their impact on IoT networks via topological sorting, and risk assessment could be achieved. Hence, to demonstrate the feasibility of the proposed method, experimental results with various scenarios has been presented, in which the security aspects have been quantified based on the network configurations. The results indicate the impact of the aspects on each other and how they could be utilized to mitigate and/or eliminate the security and privacy deficiencies in IoT networks.

## 1 Introduction

IoT is an industry, introducing smart environments where devices have capabilities such as performing tasks autonomously, making decisions via reasoning capabilities, negotiating, understanding, adapting to the environment, extracting patterns from the environment or even learning from other “things” (Sundmaekeret al., 2010). Despite the fact that IoT is among the top ten critical technology trends impacting Information Technology (IT) based on Gartner’s statement (High, 2018), it was actually invented in 1999. Kevin Ashton, cofounder of the Auto-ID Center at the Massachusetts Institute of Technology (MIT), invented the term IoT. This concept could be viewed as a convergence of three different visions named "Things" oriented, "Internet" oriented, and "Semantic" oriented (Atzori et al., 2010). In this paradigm, a variety of connected items generate data that storing, representing, searching, analyzing, and organizing them requires semantic technologies. IoT is defined by the IoT European Research Cluster (IERC) (Vermesan and Friess, 2014) as:

**Definition 1.** "*A dynamic global network infrastructure with self-configuring capabilities based on standard and interoperable communication protocols where physical and virtual "things" have identities, physical attributes, and virtual personalities and use intelligent interfaces, and are seamlessly integrated into the information network.*"

We have entered a new era that novel concepts like IoT are playing an important role in our daily lives and its significance is growing rapidly and vastly. As a result, Information and Communications Technology (ICT) players such as Amazon, Google, Apple, and Cisco are playing an important role in the IoT industry. It has been noted by (Dienna et al., 2021) that by 2025, the number of connected devices will exceed 30 billion, whereas, (Gue et al., 2021) estimates over 60 billion by 2025 with the potential to reach 125 billion devices in 2030. With this number of devices, issues such as security, data blast, scalability, and heterogeneity rise, which require close attention and deep inspection. Due to this fact, many studies have focused on this technology from various points of views. In this section, the authors have divided these research areas into four groups named architecture and technologies, applications, issues, and other.

**Architecture and technologies:** To gain knowledge about IoT, its architecture and the related technologies would be a good starting point. China's National 973 Program has supported the research on IoT architecture, and aside from introducing this technology, (Ma, 2011) has indicated a platform designed based on the theoretical studies for monitoring carbon balance as part of the GreenOrbs[1] project. On the other hand, a service oriented architecture has been presented in (Da Xu et al., 2018; Bandyopadhyay and Sen, 2011) where the latter is specifically for industries and also contains more detail about the layer interactions. In another study (Whitmore et al., 2015), architectures have been classified into hardware, software, process, and general. These groups respectively focus on the required distributed computing environment, sharing and providing access to the services, structuring business processes, and conceptual design or assessment. While healthcare specific architectures and platforms have been introduced in (Islam et al., 2015). Furthermore, (Sha et al., 2018) has presented three architectural security designs named end-to-end, edge, and distributed. In end-to-end security, the end devices are required to support IPv6 or 6LoMPAN that demands devices with rich resources. In the second type of architecture, security management tasks might be transferred from low capable end devices to more powerful edge devices. Finally, in the distributed form, an end device starts trusting the edge device after exchanging credentials between the cloud, edge and end device. On the other hand, (Jing et al., 2014) has presented a security architecture indicating various security types in each layer. Some of the existing issues in these studies are interoperability of heterogeneous systems, designing lightweight protocols and efficient layers in the architecture due to limited resources. Furthermore, authors in (Eghmazi et al., 2024) have proposed Blockchain as a Service (BaaS) based on Hyperledger Fabric, and introduced, developed, and deployed a new architecture based on it, that despite being effective in security, due to the resource usage increase in high volumes of users, it is not applicable in variety of IoT cases.

Some researchers have introduced the usage of Federated learning (FL), a distributed learning, in IoT (Khan et al., 2024; Mengistu et al. 2024). FL was initially proposed by Google, with the aim to apply the model to an immense amount of data in mobile devices (Mengistu et al. 2024). Despite being scalable and applicable for preserving privacy to some extent, there are a few challenges that reside, where resource limitation, the heterogenous nature, and data imbalance for the edge nodes is among the significant ones (Mengistu et al. 2024). (Khan et al., 2024), specifically focuses on IoMT, Smart Healthcare Systems (SHS), and enhancing the security and privacy. They proposed a privacy-preserving FL-based IDS model entitled Fed-Inforce-Fusion to identify cyberattacks in IoMT, that proves to be robust and reliable.

---

[1] http://www.greenorbs.org

IoT is related to other technologies such as cloud computing and big data. These related technologies have been inspected from various dimensions in (Lashkaripour, 2020; Lashkaripour & Balouchzahi, 2020; Da Xu et al., 2018; Lashkaripour, 2017; Islam et al., 2015; Whitmore et al., 2015; Borgia, 2014; Bandyopadhyay and Sen, 2011; Atzori et al., 2010; Bassi and Horn, 2008). Last but not least, a survey has been given in (Granjal et al., 2015) on the protocols of IoT's layered architecture with the limitations and suggestions for securing them.

**Applications:** One of the reasons that IoT is a trend, is its vast and various applications. Some of these applications have been given in (Omrany et al., 2024; Telo, 2023; Li et al., 2023; Vegesna, 2023; Abbasi et al., 2019; Atzori et al., 2010; Bandyopadhyay and Sen, 2011; Bassi and Horn, 2008; Borgia, 2014; Bujari et al., 2018; Da Xu et al., 2018; Islam et al., 2015; Whitmore et al., 2015). Due to the importance of healthcare, some studies have specifically focused on this usage. For instance, (Moosavi et al., 2016) has proposed an end-to-end security scheme for mobility enabled healthcare systems. This end-to-end security is provided by certificate-based Datagram Transport Layer Security (DTLS) handshake between end-users and smart gateways and session resumption so that the end-user and the medical sensor directly communicate. Other specifications of this promising solution are preserving confidentiality, integrity, and scalability. Another work in (Mano et al., 2016) has used a combination of image and emotions via the usage of IoT to aid patients in smart homes. An industrial version of IoT integrated with cloud technology has been proposed in (Hossain and Muhammad, 2016) for monitoring health. It has used watermarking and other related analytics to prevent identity theft and maintain a secure monitoring system. Authors in (Santos et al., 2016) have proposed an IoT-based mobile gateway solution to collect patient's location and health data so that the caretaker could act as needed. Finally, (Ahmed et al., 2024) has presented a survey of various references using IoT in the Medical section and proposed the concept of Internet of Medical Things (IoMT) and authors in (Omrany et al., 2024) specifically concentrate on researches conducted in the field of smart cities between 2010 to 2022.

**Issues:** Security has always been an important issue in various contexts including IoT. As a result, all studies in (Dhar et al, 2024; Mishra et al., 2023; Telo, 2023; Cook et al., 2023; Atzori et al., 2010; Bandyopadhyay and Sen, 2011; Bassi and Horn, 2008; Borgia, 2014; Bujari et al., 2018; Conti et al., 2018; Da Xu et al., 2018; Islam et al., 2015; Sha et al., 2018; Whitmore et al., 2015) have considered this issue either specifically or with other related aspects in IoT. A taxonomy of security threats alongside the open issues have been given in (Alaba et al., 2017; Dabbagh and Rayes, 2017; Jing et al., 2014). They have also discussed the possible attacks and countermeasures similar to (Aziz Al Kabir et al., 2023; Vegesna, 2023; Ning et al., 2013; Zhao and Ge, 2013). Furthermore, (Borgohain et al., 2015) classifies the IoT attacks based on connectivity technologies, while (Andrea et al., 2015; Hossain et al., 2015) introduce various groups of attacks and some guidelines to counter them. Moreover, four groups of attacks named physical, network, software, and encryption are presented in (Andrea et al., 2015), whereas (Hossain et al., 2015) classifies them into five layers (physical, media access control, network, transport, and application). On the other hand, whilst (Telo, 2023) presents encryption, access control, user awareness, network segmentation, and security patches and constant firmware updates as solutions to mitigate the security risks, (Alwahedi et al., 2024; Pakrooh et al., 2024; Cook et al., 2023;

Mazhar et al., 2023) demonstrate the strengths and weaknesses via comparing various techniques of Artificial Intelligence, specifically Machine Learning. (Cook et al., 2023) also considers Blockchain, while (Pakrooh et al., 2024 and Mazhar et al., 2023) presents Deep learning in this regard. Furthermore, authors in (Dhar et al, 2024) have introduced blockchain and quantum cryptography as a means to preserve the security and privacy of the multimedia data obtained from IoT networks, which requires refinement in areas such as scalability, energy efficiency, and privacy enhancement.

With the focus on Industrial IoT (IIoT), authors in (Mishra et al., 2023) have presented a comprehensive layer-based classification of attacks and mitigation techniques. Moreover, for preserving the security in IIoTs, they have demonstrated cryptographic techniques to maintain security. Last but not least, (Yan et al., 2014) has given a survey on trust management, and (Kamalov et al., 2023) presents various approaches used in IoT environments to preserve security. This paper also demonstrates the open issues, and finally proposes a research model for holistic trust management.

**Other:** The researchers have considered a separate group for the remaining inspected papers that do not fit in any of the previous groups. (Botta et al., 2016; Díaz et al., 2016) have worked on integration of cloud and IoT, and presented various features such as architecture, security, policies, and applications. Furthermore, they have introduced an intelligent collaborative security model to minimize security risks. A taxonomy of the connected objects in an IoT network is given in (Dorsemaine et al., 2015). Moreover, (Baho & Abawajy, 2023) present a survey of various studies performed in this field with thorough examination on the quality and quantity of the works, whereas the survey in (Alhilali & Montazerolghaem, 2023) focuses on load-balancing via utilizing Artificial Intelligence (AI) in Software-Defined Networks (SDNs) that can be utilized in IoT to elevate the Quality of Service (QoS) (Montazerolghaem, 2021; Montazerolghaem & Yaghmaee, 2020). At last, the building blocks of an IoT business model framework is given in (Dijkman et al., 2015) that could be a good starting point for creating business models in various IoT applications.

Due to the significance of IoT security, the authors have focused on this matter. As mentioned earlier, the studies related to security have mostly presented a classification of attacks, and open issues, whereas some gaps exist in between, such as gaining knowledge about the current network situation via a thorough list of the related security aspects, their relations, causes, and impacts. These gaps could be filled with the results of this research.

The main contributions of this work are summarized as follows:

- Introducing a comprehensive list of the security aspects in IoT.
- Classifying the aspects into five categories named data, access control, standard, network, and loss for better security management.
- Formally defining the new concept of BSAG which contains two types of nodes and three types of edges.
- Proposing a novel Bayesian dependency graph, BSAGIoT, for illustrating the relations between the proposed security aspects in any given IoT network.
- Indicating that BSAGIoT, which is a generic model applicable to any IoT network, is a DAG that could be a great reference for the security specialists to identify the cause of various security challenges, how they affect one another, and their impact on IoT networks via topological sorting.
- Demonstrating that BSAGIoT could further be used for risk assessment based on Bayes theorem.

- Presenting a case study for BSAGIoT on a sample testbed and analyzing the overall experimental results that validate the mentioned claims in different scenarios.

To the best of the authors' knowledge no such research has been done in the context of IoT so it could be of great use for security experts in tracking and management purposes. Hence the remainder of the paper is organized as follows: Section 2 introduces the related technologies and applications of IoT. While security aspects and theoretical preliminaries are introduced in the following sections respectively. Moreover, the next section presents the proposed security aspects, the related dependency rules, and BSAGIoT. The following section analyzes the proposed model via a case-study containing different scenarios to demonstrate its robustness and practical feasibility in a real-world IoT environment. Finally, the last section contains the conclusion and future works of the research.

## 2 IoT Technology and Applications

IoT networks with the vast usage in many fields are an integration of several technologies that increase the complexity of such an environment. Some of the main technologies and applications are further discussed in this section.

**Cloud Computing:** National Institute of Standards and Technology (NIST) defines cloud computing as follows (Mell and Grance, 2011):

**Definition 2.** "*Cloud computing is a model for enabling ubiquitous, convenient, on-demand network access to a shared pool of configurable computing resources (e.g., networks, servers, storage, applications, and services) that can be rapidly provisioned and released with minimal management effort or service provider interaction.*"

Cloud computing has different service models named Software as a Service (SaaS), Platform as a Service (PaaS), and Infrastructure as a Service (IaaS), and could be deployed in forms named public, private, community, hybrid, and virtual private (Zhang et al., 2010). This technology can be utilized in various crucial applications such Electronic Education Systems with the architecture introduced in (Lashkaripour & Balouchzahi, 2020) that is beneficial for all the parties involved such as the respective university or educational institute and the IT department. Although there are some security and privacy concerns in this field, as presented in (Lashkaripour 2021a & Lashkaripour 2021b), due to the increase of the IoT devices, cloud computing is of great use in issues such as scalability, availability, and reliability. In fact, cloud computing provides IoT with better capabilities in storage, management, and analysis, with a cost-efficient manner. In other words, IoT is powered by cloud's infrastructure and services due to the limited resources of its devices. For further details on the concept of cloud computing and the existing attacks and concerns please refer to (Lashkaripour 2016).

**Big Data:** The term "big data" appears to have been first used in the late 1990s by John Mashey (Dean, 2014). Among the various definitions given, the one by NIST (Chen et al., 2014) is:

**Definition 3.** "*Big data shall mean the data of which the data volume, acquisition speed, or data representation limits the capacity of using traditional relational methods to conduct effective analysis or the data which may be effectively processed with important horizontal zoom technologies.*"

Various events since 1991 with the latest being the Fifth Generation (5G) of wireless technology for digital cellular networks in 2019, have paved the way for the concept of big data as indicated in (Lashkaripour, 2020). This technology is also related to IoT, due to the vast amount of data generated by the ever-increasing connected devices that would require further action. According to (Dietrich et al., 2015), IoT is one of the sources of big data and by 2030 it would be the dominant part of big data (Marrocco et al., 2010). This huge amount of data generated by IoT devices requires big data analytics and specific tools to manipulate the data and extract the information required (Lashkaripour 2017).

**Tracking Networks:** One of the main parts in the data value chain is data acquisition, and tracking networks are vastly used for this purpose. Some of these technologies include Radio-Frequency IDentification (RFID), Near Field Communication (NFC), and Wireless Sensor Networks (WSNs) (Raggett, 2016). RFID systems contain one or more readers and several tags. This technology is used for identifying and tracking the tags attached to objects or living things that could be used in manufacturing, implanting, and recycling. NFC is a technology used for communication between devices either via contact or wirelessly. It is based on RFID but with a bidirectional characteristic. NFC can be useful for payment systems, social networking, and gaming. WSNs consist of a number of sensing nodes that are in charge of sensing a special phenomenon like light or pressure, and sending the results to the network sinks for further inspection. The range of IoT use cases and the diversity of devices inside the network, results different tracking networks for communication and data exchange, as mentioned earlier.

**Access Networks:** In an IoT network, a large group of heterogeneous devices communicate with each other that require a variety of access networks. Some of these technologies are satellite, Worldwide Interoperability for Microwave Access (WiMAX), Ethernet, Wi-Fi, cellular, Zigbee, and Bluetooth. They are known by different characteristics such as frequency bands, data rate, and maximum distance (Borgia, 2014). The usage of these access networks depends on the scope and application of IoT.

The IoT applications mentioned in this section rely on networks of gateways, smart sensors, and actuators. The advantage of this network would be, providing an easier and safer life via performing the right action at the right time (for more information please refer to (Vermesan and Friess, 2014)).

**Smart Health:** The gadgets and monitoring devices used for healthcare are increasing each day. They are useful for both patients and doctors by facilitating health support. Furthermore, IoT could also be used to identify and authenticate employees and prevent theft via object tracking. Due to the importance of health in general and the value and cost of human lives for governments, IoT could help prevent food and drug counterfeiting. Via this technology, it is possible to trace and identify the origin of the product for emergency cases.

**Smart Manufacturing***:* In this application, the state and step of the products manufacturing process, when and where the product was made, and the used substances, could be easily traced. If more detail is demanded, it is even possible to track a product through its assembly line. Aside from having a smart manufacturing environment, the interaction with outer environments like the retailers for stock and sales data, is possible and very useful.

**Smart Logistics:** Logistics management is part of the supply chain management that plans, implements, and controls the flow of sending and receiving goods and services. All this is done to meet the requirements of the customers, increase their satisfaction, and also assist manufacturers with feedback from the market. By replacing the bar codes with electronic tags, the executable code inside them could improve routing via intelligent interaction and decision making.

**Smart Grid:** Smart grid is expected to be the implementation of a kind of "Internet" in which energy is managed similarly to the data to decide the best pathway for the packet to reach its destination with the best integrity levels. Since relying on fossil resources does not have a future, and nuclear energy is not a future proof option, resources like green power should be replaced.

**Smart Transportation:** The communication of vehicles, via internet could be a great help to traffic management through safe and fast transportation. Furthermore, vehicles would need communication with the grid, infrastructures, vehicles, and devices to maintain themselves, manage energy consumption, congestion, billing, and traffic. This type of network is very useful for driverless vehicles that demand such information.

**Smart Buildings:** In a smart building we have monitoring and controlling equipment, and the state of the building that would contain heating, cooling, air conditioning, lighting, security, parking, and waste. Robots could use the sensors for housekeeping activities and this could be useful for the disabled and elderly. It would also lead to safe and secure buildings, better resource management, and cost efficiency. For example, it is possible to be informed whether your belongings have been moved elsewhere without permission, and also track them.

**Smart Maritimes:** In a smart maritime, various departments can cooperate in real-time, which would lead to better performance and efficiency in finance. Part of this could contain monitoring shipboard equipment and machinery, in addition to opting routes that are fuel-efficient. Since 55% of the ship's operating cost is related to the fuel consumption, IoT could play a significant role in not only cutting off costs but also minimizing CO2 emissions (Gerodimos et al., 2023), which at this point has a tremendous effect on global warming.

## 3 Security Aspects in IoT

Despite all the benefits that a novel paradigm presents there are always some issues that require specific attention. A variety of challenges such as security, energy consumption, standardization, heterogeneity, and mobility exist. Due to the significance of security in IoT, where different users share a variety of data with a high frequency, the researchers have focused on this matter. The security concerns, briefly discussed in this section, vary depending on the user that could be a consumer, enterprise, and/or government.

### *3.1 Security standards and policies*

Standards and policies are the only way to secure the IoT network, and other aspects are a subset of this main one. As a result, we should take advantage of the available resources, for instance, Open Web Application Security Project (OWASP) IoT Security Guidance, and World Wide Web Consortium (W3C) security activity (Raggett, 2016). Moreover, policymakers have significant opportunities to create spaces for exploring challenges and identifying solutions (Abendroth et al., 2017).

### *3.2 Prohibition laws and regulations*

The regulations by which government, police, and intelligence agencies get access to data, differ in various regions of the world. Since these laws and regulation are not as rich as they should be due to the novelty of IoT technology, improvement and evolution is required in this field.

### *3.3 Privacy, Trust, and related policies*

In an IoT network, facilities might be delivered by multiple providers with different security and privacy policies that integrating them without any suitable access control level could lead to a security breach. Hence, an effective communication among the devices requires a trust framework to manage their interaction and sharing. Furthermore, the outsourced data has the potential risk of unauthorized access that could lead to privacy violation. Accordingly, one of the ways that maintain individual's privacy is interaction with trusted devices.

### *3.4 Compliance issues*

The variety of technologies and protocols used in an IoT network and also the additional layers, increase the compliance issues (Regan et al., 2016). Aside from this, IoT security controls might interfere with personal expectation of privacy. Furthermore, legislation difference in geographical regions can also increase the compliance issues.

### *3.5 Quality of Service (QoS)*

QoS contains a variety of factors such as reliability, efficiency, functional stability, and load balance (Montazerolghaem & Yaghmaee, 2020) that help present a better service to users. Each layer of an IoT architecture has its own QoS factors, and among them communication layers (physical, link, and network) are highly important (White et al., 2017).

### *3.6 Security misconfiguration*

It can happen at any level of an application stack, including the platform, web server, application server, framework, and custom code (Khalil et al., 2014). Since recovery costs could be very expensive, developers and network administrators need to work together to ensure that the entire stack is configured properly.

### *3.7 Access Control*

Fine-grained access control policies are essential due to heterogeneity, diversity of protocols, and access requirements. Furthermore, generic access control interfaces are needed for interoperability but it should be considered that the least privilege policy is of importance. Ragothaman and colleagues in (Ragothaman et al., 2023) present a variety of Access Control (AC) models such as Role-Based Access Control (RBAC), Organization-Based Access Control (OrBAC), Usage-Based Access Control (UCON), Organization-Based Access Control (OrBAC), and Blockchain-Based Access Control (BBAC).

### *3.8 Authentication and Authorization*

Authentication is the process of determining whether someone or something is who or what it claims to be. This aspect is a central element to address the privacy and security issues of an IoT network. Furthermore, specifying the access rights to resources either physical or not, is known as authorization. Any resource, can be easily accessible for authorized users such as humans, machines, services or network objects.

### *3.9 Confidentiality, Integrity, and Availability (CIA)*

CIA triad is known as the building block of a secure system (Zissis and Lekkas, 2012). Confidentiality is confirmed only when authorized users have access to the network. Lack of strong authentication could lead to unauthorized access that affects privacy. Integrity on the other hand means, preserving the accuracy and completion of the assets in a way that they can only be modified by authorized parties or in authorized ways. Finally, availability is referred to an authorized entity being able to access the data, software and also hardware whenever required. System availability is the system's ability to carry on operations even on the misbehave of some authorities or a security breach.

## 4 Theoretical Preliminaries

In this section, the authors present the preliminaries required for understanding BSAGIoT in the next section. This information is also used for the analysis and risk assessment of BSAGIoT, and for indicating how a security specialist could use it to reduce the risks in an IoT network. For this purpose, Bayesian network and the new concept of BSAG are explained in detail in the following.

### *4.1 Bayesian Network*

A Bayesian network is a DAG indicating the joint probability distribution of a set of random variables, where the nodes are the random variables and the edges are static causal relationship of the related nodes. Each node has a conditional probability distribution that quantifies the impact of parent nodes (Horný, 2014). If X = {x1, x2, …, xn} is the set of random variables inside the DAG, the joint probability distribution of X is (Munoz-González et al., 2017):

$$P(X) = \prod_{i=1}^{n} P(xi|Pa(xi)) \quad (1)$$

Where Pa(xi) also shown as Pai denotes the parents of node xi.
Bayesian network is based on Bayes theorem, where the conditional probability of X given Y is:

$$P(X|Y) = P(X)P(Y|X)/P(Y), \quad P(Y) \neq 0 \quad (2)$$

In this equation, P(X|Y) is the posterior probability, P(X) is the prior probability, P(Y|X) is the probability of observing Y given the cause X, and P(Y) is unconditional probability of Y.

### *4.2 Probability Calculation*

IoT networks, composed of numerous interconnected and vulnerable components, present a significant security challenge. The sheer volume of potential vulnerabilities makes accurately assessing the probability of successful exploitation based on expert knowledge extremely difficult. To calculate the exploitation probability of each vulnerability in the IoT network, which will then be used as input for a Bayesian network, we utilize the Common Vulnerability Scoring System (CVSS) metrics (Mell et al., 2007).

To quantitatively assess the severity of security vulnerabilities, the Common Vulnerability Scoring System (CVSS) utilizes three metric groups: Base, Temporal, and Environmental. Security experts can tailor these metrics based on specific network characteristics to generate a vulnerability score between 0 and 10. This paper focuses on the CVSS Base Score, which reflects the inherent characteristics of a vulnerability. Calculating the CVSS version 3.0 Base Score requires initializing several key metrics, including Attack Vector (AV), Attack Complexity (AC), Privileges Required (PR), User Interaction (UI), Scope (S), Confidentiality Impact (C), Integrity Impact (I), and Availability Impact (A). In summary, AV reflects the context in which a vulnerability can be exploited. AC describes the conditions that must be present for successful exploitation. PR indicates the level of access needed to exploit the vulnerability. UI captures the requirement for a normal user to participate in the vulnerability exploitation. S identifies the potential for a vulnerability in one software component to impact other components. Finally, C, I, and A measure the impact on the CIA triad of the information resources resulting from a successful exploit. After initializing the metrics required by the CVSS version 3.0 calculator (CVSS Version 3.0 Calculator, 2025), security experts can calculate the Base Score for each vulnerability. These metric values are often readily available for individual CVE vulnerabilities in existing vulnerability databases, such as the National Vulnerability Database (NVD) (NIST-NVD, 2025).

### *4.3 Bayesian Security Aspect Graph*

Bayesian networks are suitable to model the proposed Security Aspect Graph (SAG). A logical SAG is defined as a bipartite graph representing dependencies between security aspects, either known as states or vulnerabilities. This graph is formally defined by the authors, as:

**Definition 4.** *A directed bipartite graph G = (S ∪ V, Ri ∪ Rr), where the vertices are the sets of states (S) and vulnerabilities (V), and the edges are relations named imply (Ri ⊆ S × V), result (Rr ⊆ V × S), and lead ($R_l$ ⊆ V × V). That is, the states imply the vulnerabilities, the vulnerabilities result the states, and the vulnerabilities lead to vulnerabilities.*

To be more precise, we can define the vulnerability and state as:

**Definition 5.** *A vulnerability is a weakness within the hardware and/or software component(s) of a network due to design, implementation, deployment, and/or configuration deficiencies. The vulnerability could be lead from another vulnerability in the network, or result a state.*

**Definition 6.** *A state is a condition or property of a network that is an implication of a vulnerability.*

According to Definitions 4-6, a vulnerability is either lead from another vulnerability or the state implies the vulnerability, and a state is a result of a vulnerability. Thus, a SAG indicates the paths that an attacker can take to exploit a set of vulnerabilities and pass some states in the network to achieve his goal. As a result, BSAG could be used for further analysis and calculating the probability that an attacker can reach each node in the graph. In a BSAG the edges entering a node could have two types of relations either AND or Or. Hence, the conditional probability of a node given the states of its parents and the given relation could be computed based on the below equations:

- AND (Munoz-González et al., 2017): All the preconditions should be met to compromise node xi (xj = F: xj is not compromised).

$$P(xi|Pai) = \begin{cases} 0 & \exists\, xj \in Pai \mid xj = F \\ \prod_{j:xj} Pvj & otherwise \end{cases} \quad (3)$$

- OR (Liu and Man, 2005): At least one precondition needs to be satisfied to compromise node xi.

$$P(xi|Pai) = \begin{cases} 0 & \forall\, xj \in Pai \mid xj = F \\ 1 - \prod_{j:xj} (1 - Pvj) & otherwise \end{cases} \quad (4)$$

Pvj used in equations (3) and (4) is the probability of an attacker successfully exploiting a vulnerability vj. which can be calculated according to Section 4.2. These quantified values are then used to propagate the probabilities and calculate the probability of the remaining nodes in the BSAG via equations (1) to (4).

## 5 Proposed Security Aspects and BSAGIoT

In this section, the comprehensive list of proposed security aspects in an IoT network, that are either the result of a thorough analysis by the authors on the various cited references or extracted from the authors' work and experience in the context of IoT security, are introduced, and the dependency graph that shows how these aspects lead to one another is demonstrated. In other words, the relation between the security aspects is presented in the form of rules and illustrated by the dependency graph titled BSAGIoT. This dependency graph is also a BSAG (Section 4.2), a novel concept introduced by the authors. To the best of the authors' knowledge no such work has been presented for this evolving paradigm.
In addition, the security expert has the ability to assess the power of the potential attackers in any given IoT network of any application via quantifying the specifications of the IoT network and computing the conditional probability of exploiting the security aspects in BSAGIoT. Furthermore, the impact of the introduced aspects on an IoT network could be easily concluded from the graph. Hence, this paper could be a great reference for assisting IoT security experts, as BSAGIoT is a generic model applicable to any IoT network.

### *5.1 Proposed Security Aspects*

Securing an IoT network demands close attention. Hence, the authors present the concluded security aspects, which have stemmed from the thorough analysis on an extensive set of books and papers in the field of IoT security or the authors' works and experience in this context, and also demonstrate the dependencies between these security aspects. These dependencies are indicated in the form of rules as presented in Table 1. Each dependency rule is given a number along with the references used in concluding such a rule. Those rules with no reference are the results of the authors' works and experience in the context of IoT security. This table contains the rule numbers (R1 ~ R29), the components inside a dependency rule, and the reference(s) to back up the rule respectively. A dependency rule is presented in the form of A → B, where A, referred to as the Left-Hand Side (LHS), is known as antecedent or premise and B, referred to as the Right-Hand Side (RHS), is known as consequent or conclusion. These twenty nine rules assist specialists in protecting any IoT network against various existing threats, vulnerabilities, and attacks.

**Table 1** Proposed dependency rules between security aspects of IoT

| NO | RULE | REFERENCES |
|---|---|---|
| 1 | Data confidentiality, integrity, and/or availability breach (A2) → QoS violation (A1) | |
| 2 | Data alteration, inconsistency, and/or loss (A3) → Data confidentiality, integrity, and/or availability breach (A2) | (Ahmed et al., 2024; Ning et al., 2013; Atzori et al., 2010; Medaglia and Serbanati, 2010) |
| 3 | Data privacy violation (A4) → Data alteration, inconsistency, and/or loss (A3) | (Jing et al., 2014; Atzori et al., 2010) |
| 4 | Data leakage (A5) → Data privacy violation (A4) | (Telo, 2023; Jing et al., 2014) |
| 5 | Identity theft or forging legitimate user credentials (A6) → Data leakage (A5) | (Ahmed et al., 2024; Telo, 2023; Ahmadvand et al., 2023; Allouzi et al. , 2021) |
| 6 | Plain text traffic (A7) → Data leakage (A5) | (Baho & Abawajy, 2023; Khurshid et al., 2023; Allouzi et al., 2021) |
| 7 | Identity theft or forging legitimate user credentials (A6) → Financial loss (A8), Blackmail or fraud (A9) | (Jahangeer et al., 2023; Telo, 2023) |
| 8 | Privacy and trust violation (A10) → Identity theft or forging legitimate user credentials (A6) | (Telo, 2023; Atzori et al., 2010; Bujari et al., 2018; Sha et al., 2018) |
| 9 | Public data misuse (A11) → Privacy and trust violation (A10) | (Cook et al., 2023) |
| 10 | Authentication and access control flaw (A12) → Privacy and trust violation (A10) | (Mengistu et al. 2024; Telo, 2023; Ragothaman et al., 2023; Allouzi et al., 2021; Botta et al., 2016; Díaz et al., 2016; Jing et al., 2014; Sicari et al., 2015; Vermesan and Friess, 2014; Yan et al., 2014; Atzori et al., 2010) |
| 11 | Insufficient authorization (A13) → Privacy and trust violation (A10) | (Allouzi et al. , 2021; Botta, et al., 2016; Díaz et al., 2016; Sicari, et al., 2015; Yan et al., 2014; Atzori et al., 2010) |
| 12 | Malicious nodes (A14) → Privacy and trust violation (A10) | (Jahangeer et al., 2023; Telo, 2023; Allouzi et al., 2021; Andrea et al., 2015; Borgohain et al., 2015) |
| 13 | Service disrupt (A16) → Health and/or life(s) at risk (A15) | (Pakrooh et al., 2024; Telo, 2023) |
| 14 | Malicious nodes (A14) → Service disrupt (A16) | (Telo, 2023) |
| 15 | Credential disclosure (A17) → Authentication and access control flaw (A12) | (Conti et al., 2018; Yan et al., 2014) |
| 16 | Node hijacking (A18) → Malicious nodes (A14), Credential disclosure (A17) | (Andrea et al., 2015; Borgohain et al., 2015; Zhao and Ge, 2013) |
| 17 | Hardware and/or software compromise (A19) → Node hijacking (A18) | (Mengistu et al. 2024; Telo, 2023; Ahmadvand et al., 2023; Kabir, 2023; Sha et al., 2018) |
| 18 | Insecure network (A20) → Hardware and/or software compromise (A19) | (Baho & Abawajy, 2023; Telo, 2023; Atzori et al., 2010; Jing et al., 2014; Sha et al., 2018; Jing et al., 2014) |
| 19 | Security misconfiguration (A21) → Insecure network (A20) | |
| 20 | Compliance issues (A22) → Node hijacking (A18) | (Khalil et al., 2014) |
| 21 | Lack of regular firmware updates or patch installations (A23) → | (Ahmed et al., 2024; Telo, 2023; Cook et al., 2023; Khurshid et al., |
| | Insecure network (A20) | 2023; Sadhu et al. 2022) |
| 22 | Lack of security standards and policies (A24) → Compliance issues (A22) | (Ahmed et al., 2024; Mengistu et al. 2024; Cook et al., 2023; Gerodimos, 2023; Mahmoud et al., 2015) |
| 23 | Lack of security standards and policies (A24) → Insecure network (A20), Hardware and/or software compromise (A19), Authentication and access control flaw (A12) | (Mahmoud et al., 2015) |
| 24 | Track Nodes (A26) → Privacy and trust violation (A10) | |
| 25 | Insecure interfaces (A25) → Track Nodes (A26) | (Allouzi et al. , 2021) |
| 26 | Insecure interfaces (A25) → Authentication and access control flaw (A12), Node hijacking (A18) | |
| 27 | Lack of account lockout (A27), Weak credentials (A28) → Insecure interfaces (A25) | (Allouzi et al. , 2021) |
| 28 | Lack of prohibition laws and regulations (A29) → Privacy and trust violation (A10) | (Telo, 2023; Cook et al., 2023; Weber, 2010; Yan et al., 2014) |
| 29 | Application and networking protocols deviation (A30) → Insecure network (A20), Privacy and trust violation (A10) | (Yan et al., 2014) |

A brief analysis of the overall proposed rules containing the introduced aspects (Ai) is given as follows:

With all the convenience that IoT technology conveys, it can also be a threat to the privacy of its users in various applications. Hence, authentication, authorization roles, and policies become necessary to control the access to any resource especially sensitive data (Botta et al., 2016). On the other hand, since privacy and trust are interconnected, preserving privacy would build trust among the users (consumers, enterprises, and/or governments), and trust is an essential building block of IoT communication and data exchange. The heterogeneous nature of this dynamic network of things, leads to a variety of technologies and protocols that could cause compliance issues and protocol deviations. Therefore, it is inevitable to have a secure network and preserve the demanded privacy and trust without considering these factors (Yan et al., 2014).

Another important side to IoT environments is the related standards, policies, and regulations. This part also requires careful attention due to their prohibition role against violations, compliance issues, and insecurity (Ahmed et al., 2024; Cook et al., 2023; Gerodimos, 2023; Mahmoud et al., 2015; Weber, 2010). It should be mentioned that insecurity includes network level, application level, interfaces, and respective misconfigurations; therefore their significance is undeniable. The result of such breaches would be hardware and/or software compromise, or even node hijack that could further lead to malicious nodes with destructive intentions (Telo, 2023; Andrea et al., 2015). A malicious node could modify data before, during or after transmission, or seek to deny or disrupt service to other nodes that could put health and/or life(s) at risk. Therefore, in addition to privacy and trust violation, the consumed data may not reflect the most recent updates.

As it is demonstrated in Table 1, all the security aspects eventually lead to one or more of the aspects entitled QoS violation (A1), Financial loss (A8), Blackmail or fraud (A9), and/or Health and/or life(s) at risk (A15), which are a direct result of Data confidentiality, integrity, and/or availability breach (A2), Identity theft or forging legitimate user credentials (A6), and service disrupt (A16), respectively. Moreover, users of the IoT network demand services that preserve their personal sensitive data. Therefore, data as one of the main assets in any organization, requires high protection against alteration, inconsistency, and/or loss that would preserve the CIA triad, and meet the QoS requirements. It is well known that a network is only as strong as its weakest node or link. Therefore, considering all the mentioned security aspects for each and every component of the IoT environment is essential.

The proposed security aspects in Table 1, are shown as Ai, where A comes from the word Aspect and i is the respective number. It should also be stated that all the aspects, regardless of the rule, are either a State (S) or a Vulnerability (V), where the states imply the vulnerabilities, and the vulnerabilities result the states or lead to other vulnerabilities.

The step-by-step approach followed to extract the resulted rules is summarized below:

- Initially, a vast number of books, journal articles, and conference papers in the field of IoT security were selected based on factors such as citation volume and direct relevance to the objective of the current study. In total, 94 sources published between 2005 and 2024 were selected for in-depth examination and analysis.
- Since the goal was to develop a generic model applicable to any IoT network, the authors prioritized extracting claims, statements, analysis results, and conclusions specifically relevant to IoT networks in a broad and general context from the researched literature.
- Next, with our novel proposed concept of BSAG consisting of the aspects (either a state or a vulnerability) and the three valid rule types, we aimed to determine whether the context of the reference corresponds to a vulnerability that results a state, or a vulnerability that leads to another vulnerability, or a state that implies a vulnerability.
- Hence, the identified aspects, and the dependencies between them was presented in the form of rules.
- Ultimately, the finalized rules underwent validation by a security expert and were further refined based on their feedback, if necessary.

Moreover, the high volume of rules and the related aspects demands a classification that could aid the security specialist from various perspectives. For instance, increasing the accuracy of the IoT network security inspection when narrowing down the relative aspects to a specific category under investigation. In addition, the speed in which the analysis is performed is elevated as the breadth of the investigation is alleviated to a single category. Therefore, this classification determines the breadth of the security expert’s point of view towards purposes such as precaution, protection, cause identification, and management that could easily be broadened or limited. Consequently, to simplify the overall process for the security expert, the authors have classified the security aspects given in Table 1, into five categories named data, access control, standard, network, and loss. Table 2 describes each category and specifies the related aspects that are relative to it among the total thirty aspects introduced in the last column of this table. In addition, these classifications are used in the dependency graph for further analysis, as indicated in Section 5.2. For each category mentioned in Table 2, the researchers have given a brief definition below and aspects have been classified into the categories accordingly.

**Data:** This category covers the aspects related to data security during its lifecycle. Since data is one of the main assets in any organization its maintenance and protection is of great importance. Issues such as alteration, inconsistency, loss, privacy violation, and leakage, violate at least one of the CIA triad.

**Table 2** Category description and the security aspects related to each category

| CATEGORY | NAME | DESCRIPTION | RELATED SECURITY ASPECTS |
|---|---|---|---|
| **C1** | Data | The building blocks of data security aspects including CIA. | A2: Confidentiality, integrity, and/or availability.<br>A3: Alteration, inconsistency, and/or loss.<br>A4: Privacy violation.<br>A5: Leakage.<br>A7: Plain text traffic.<br>A11: Public data misuse.<br>A26: Track Nodes. |
| **C2** | Access control | The aspects related to authentication and authorization to control the level of access rights to various resources. | A6: Identity theft or forging legitimate user credentials.<br>A12: Authentication and access control flaw.<br>A13: Insufficient authorization.<br>A14: Malicious nodes.<br>A17: Credential disclosure.<br>A18: Node hijacking.<br>A19: Hardware and/or software compromise.<br>A27: Lack of account lockout.<br>A28: Weak credentials. |
| **C3** | Standard | The standards required for securing the environment and its users from diverse attacks while at the same time reliability, flexibility, and performance is preserved. | A10: Privacy and trust violation.<br>A22: Compliance issues.<br>A23: Lack of regular firmware updates or patch installations.<br>A24: Lack of security standards and policies.<br>A29: Lack of prohibition laws and regulations. |
| **C4** | Network | The network used for communication and data exchange between numerous devices with various technologies, requires maintenance and preserving QoS factors and security. | A1: QoS violation.<br>A16: Service disrupt.<br>A20: Insecure network.<br>A21: Security misconfiguration.<br>A25: Insecure interfaces.<br>A30: Application and networking protocols deviation. |
| **C5** | Loss | The aspects that could lead to any form of loss, from extortion to more severe consequences that impacts human health and/or life(s). | A8: Financial loss<br>A9: Blackmail or fraud<br>A15: Health and/or life(s) at risk |

**Access control:** It covers the aspects related to controlling the access to various resources either physical or non-physical as indicated in the previous section. For this purpose, users should be authenticated and authorized according to their accessibility level.

**Standard:** The standard category covers the aspects related to regulatory and governing authorities that define the policies required for preserving security and preventing various attacks by taking precaution measures. These policies are related to various parts in an IoT network. For example, a significant part of these aspects is related to cloud computing.

**Network:** The fourth category covers the aspects related to the vast IoT network that not only the number of its interconnected devices is increasing, but also the amount of transferred data is widely extending. In this category, network specific attacks such as Denial of Service (DoS), Distributed DoS (DDoS), flooding attack, and internet protocol vulnerabilities require special attention.

**Loss:** The final category consists of the aspects related to any form of loss. This could contain extortion and/or blackmail or any other form of financial loss, in addition to more severe consequences that could impact the health and lives of individuals with more critical outcomes that might not be redeemable at any cost.

### *5.2 BSAGIoT*

Figure 1 demonstrates the proposed generic model named BSAGIoT, that could be applied to any IoT network. This novel dependency graph proposed, is the result of thorough inspection in the field of IoT security via utilizing various resources in this context. The overall structure of the extracted dependencies among the proposed aspects is as illustrated in Figure 1, where the nodes (aspects) and the edges (rules) are the building blocks of BSAGIoT.

As a result, in a generic rule of Ai → Aj, we can have any of the following conditions, where:

- V → S: The vulnerability results the state.
- V → V: The vulnerability leads to another vulnerability.
- S → V: The state implies the vulnerability.

To further elaborate on the rules presented in Table 1, please find the following examples as to where we have either a vulnerability or a state on the LHS or the RHS of a rule and how it is structured.

V → S: It is worth noting that whenever a vulnerability is exploited the state in which that vulnerability is compromised is resulted. Hence, for rule number 9 (R9) as an example, the vulnerability on the LHS when exploited results the available public data being misused. A real-life example of this state has been mention in (Vigdor, 2022), where a teenager with his Twitter account used the publicly available data, mandated by United States Federal Aviation Administration (FAA) (ADS-B Rules; 2025), transmitted from Elon Musk's private jet's transponders, responsible for recording the flight's location. This data was gathered through a service known as ADS-B Exchange which collects unfiltered flight. (Cook et al., 2023) claims that this public data that would seem to be of little value has been misuse and can be personally harmful.

V → V: Rule number 4 (R4) as an example.

*Data leakage (A5) → Data privacy violation (A4)* *(Telo, 2023; Jing et al., 2014)*

In this case, we have both A5 and A4 as a vulnerability, where A5 leads to A4. Telo mentions in multiple instances that the growing use of IoT networks has increased the frequency of sensitive data leaks, leading to privacy violations. On the other hand, authors in (Jing et al., 2014) claim that the theft of information leads to privacy exposure.

**Figure 1** BSAGIoT

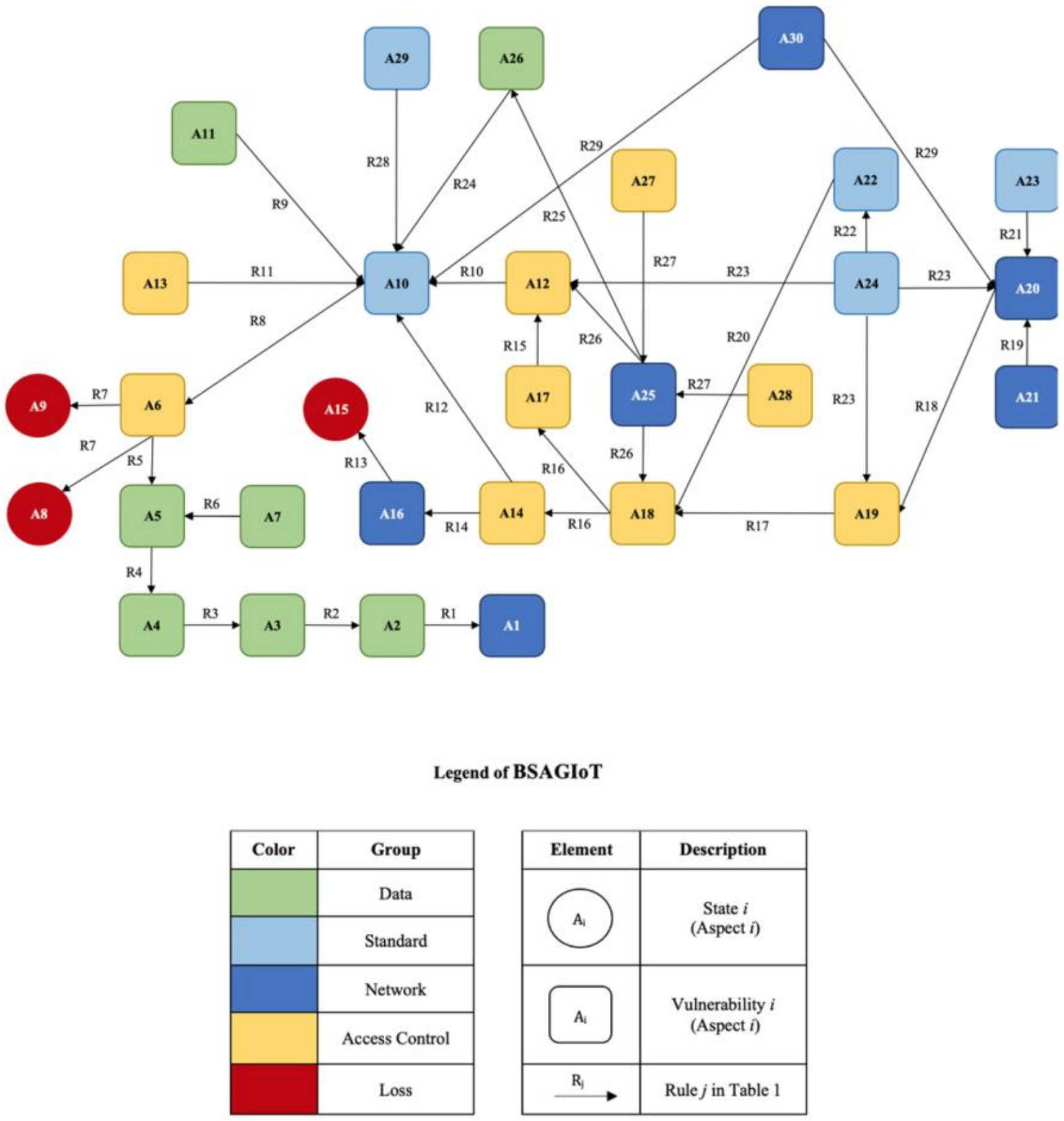

S → V: Rule number 7 (R7) as an example.

*Identity theft or forging legitimate user credentials (A6) → Financial loss (A8), Blackmail or fraud (A9)* *(Jahangeer et al., 2023; Telo, 2023)*
In this case, we have A6 as the vulnerability resulting, the state where identity theft or forging legitimate credentials' vulnerability is successfully exploited that leads to two different states of A8 and A9. In (Jahangeer et al., 2023), the authors have stated that insecure or weak networks serve as prime targets for attackers seeking to access confidential data. Through identity theft, such breaches can lead to financial losses. On the other hand, both A8 and A9 have been clearly mention in (Telo, 2023). Telo states that personal and/or sensitive data collected in an IoT network, such as health data in hospitals or financial data in smart homes, can be exploited for identity theft, extortion, or fraudulent transactions, potentially resulting blackmail or fraud if intercepted by an attacker.

To further illustrate, ten sample rules in addition to the ones mentioned earlier are provided as instances in Table 3. The sample statements were extracted from the specified references, with only one statement indicated for each rule, even though, in some cases, multiple statements were available.

**Table 3** Sample statements indicating the origin of dependency rules

| NO | A SAMPLE STATEMENT | RULE |
|---|---|---|
| R3 | User data privacy must be guaranteed because users require maximum protection for their personal information (Atzori et al., 2010). | Data privacy violation (A4) → Data alteration, inconsistency, and/or loss (A3) |
| R11 | Unauthorized system entities cannot access private data of others in data communications and transmission. This objective is related to the security and privacy properties of IoT system wherein light security/trust/privacy solution is needed (Yan et al., 2014). | Insufficient authorization (A13) → Privacy and trust violation (A10) |
| R12 | In acknowledgement flooding, a malicious node spoofs the acknowledgements providing false information to the destined neighboring nodes (Borgohain et al., 2015). | Malicious nodes (A14) → Privacy and trust violation (A10) |
| R13 | Similarly, an attacker could exploit vulnerabilities in an IoT device used in a smart city project to gain access to critical infrastructure systems, such as traffic lights or water treatment plants. Once an attacker gains access to such systems, they can cause significant damage, disrupt services, and put lives at risk (Telo, 2023).<br>For example, if an attacker manipulates data related to traffic patterns or transportation schedules, it could lead to accidents or traffic congestion. Similarly, if an attacker injects malware into the system, it could disrupt essential services such as water or power, leading to widespread damage and disruption (Telo, 2023). | Service disrupt (A16) → Health and/or life(s) at risk (A15) |
| R16 | In node capturing, key nodes are controlled easily by the attackers such as gateway node. It may leak all information, including group communication key, radio key, matching key etc., and then threats the security of the entire network (Zhao and Ge, 2013). | Node hijacking (A18) → Malicious nodes (A14), Credential disclosure (A17) |
| R18 | Imagine a legacy physical device that uses a driver that only works with an old Operating Systems (OSs) that is no longer supported and updated by the vendor. Obviously, the old OS has to stay but it becomes a serious vulnerability. The whole system may be compromised through this weakest link. In addition, with the tight coupling of the physical system and the cyber world, compromising one can put the other at great risks and negative impact can propagate both ways. For example, compromising the cyber part of the systems allows the attackers to control the physical system (Sha et al., 2018). | Insecure network (A20) → Hardware and/or software compromise (A19) |
| R21 | Update the firmware and software on your IoMT devices, gateways, and network infrastructure on a regular basis. This protects the system from any new threats that may arise by applying the latest security updates and fixing any known vulnerabilities (Ahmed et al., 2024). | Lack of regular firmware updates or patch installations (A23) → Insecure network (A20) |
| R22 | Although there have been some successful attempts to standardize IoMT technologies, there is still a lack of universal standards that can make it difficult for different IoMT technologies to work together smoothly (Ahmed et al., 2024). | Lack of security standards and policies (A24) → Compliance issues (A22) |
| R27 | An insecure web interface can be present when issues such as lack of account lockout or weak credentials are present (Allouzi et al., 2021). | Lack of account lockout (A27), Weak credentials (A28) → Insecure interfaces (A25) |
| R28 | The establishment of an adequate legal framework for the protection | Lack of prohibition laws |
| | of security and privacy in the IoT is a phenomenon (Weber, 2010).<br>Privacy is related to trust, and a trustworthy digital system should preserve its users' privacy, which is one of the ways to gain user trust (Yan et al., 2014). | and regulations (A29) → Privacy and trust violation (A10) |

In Figure 1, aspects are presented as nodes and the relations between the aspects presented as edges. To differentiate the two aspect types from one and other, as indicated in the Legend of this figure, states and vulnerabilities have been presented as circles and squares respectively. Moreover, the relations between the aspects could be either of the three edges where the state is an implication of a vulnerability, and the vulnerability could be lead from another vulnerability in the network. It is worth noting that in this dependency graph, after each vulnerability the state in which that vulnerability has been compromised is resulted. However, to not complicate the model, the authors decided to omit these states, since they convey when the vulnerability was exploited and it can simply be specified from the aspect itself. As a result, the overall proposed rules presented in Table 1 have been modeled via the proposed novel concept of BSAG and the given two types of nodes and three types of edges. Hence, the final outcome of following the step-by-step approach mentioned in Section 5.1 and the details given in this section, is BSAGIoT. This generic model, could be of great use for precaution and protection purposes, and also for identifying the cause(s) and consequence(s) of various security challenges. For a better understanding and further transparency in the proposed aspects, each of the five categories introduced in the previous section (Table 2), is demonstrated with a specific color as illustrated in the legend of Figure 1. It should be mentioned that for convenience in the analysis and interpretation, the colors utilized for each category is consistent throughout the paper. BSAGIoT, shows the interconnection between various security aspects of the mentioned categories which leads to superior security management. It also demonstrates how aspects from different categories have impact on each other or on an IoT network.

In a BSAG, all the external nodes, containing no entry edges, also known as entry points, are vulnerabilities, while the internal nodes could be of either type. BSAGIoT is a DAG that topological sorting helps specify the root cause of a specific security aspect, how they affect one another, and also the results of such a security violation. It is worth mentioning that at least one topological sorting exists for any given DAG. In such a linear ordering, for every directed edge uv from u to v in the DAG, vertex u comes before vertex v. For any sample aspect in this linear ordering the cause(s) and consequence(s) of the aspect are respectively placed somewhere on the left hand side and the right hand side of it. For example, a direct cause of A5 (data leakage) is A6 (Identity theft or forging legitimate user credentials), and the direct consequence is A4 (Data privacy violation), which could be propagated for even deeper information. As a result, BSAGIoT could be a suitable reference for tracking and managing security when it comes to an IoT network.

As BSAGIoT illustrates and Table 2 differentiates, we have a distribution of the proposed aspects among the various categories. To comprehend the impact of the proposed categories and how they are oriented, the quantity of the security aspects in each of the five categories out of the total of thirty introduced aspects is demonstrated in Figure 2. In this figure, the order of frequency for the categories from low to high is, loss, standard, network, data, and access control. According to the researches and studies performed in this area, this ranking demonstrates where the concerns, limitations, and vulnerabilities exist. Hence, to achieve the target of mitigating and/or eliminating the security and privacy deficiencies in IoT networks, further work is required.

**Figure 2** Distribution of security aspects based on the category type

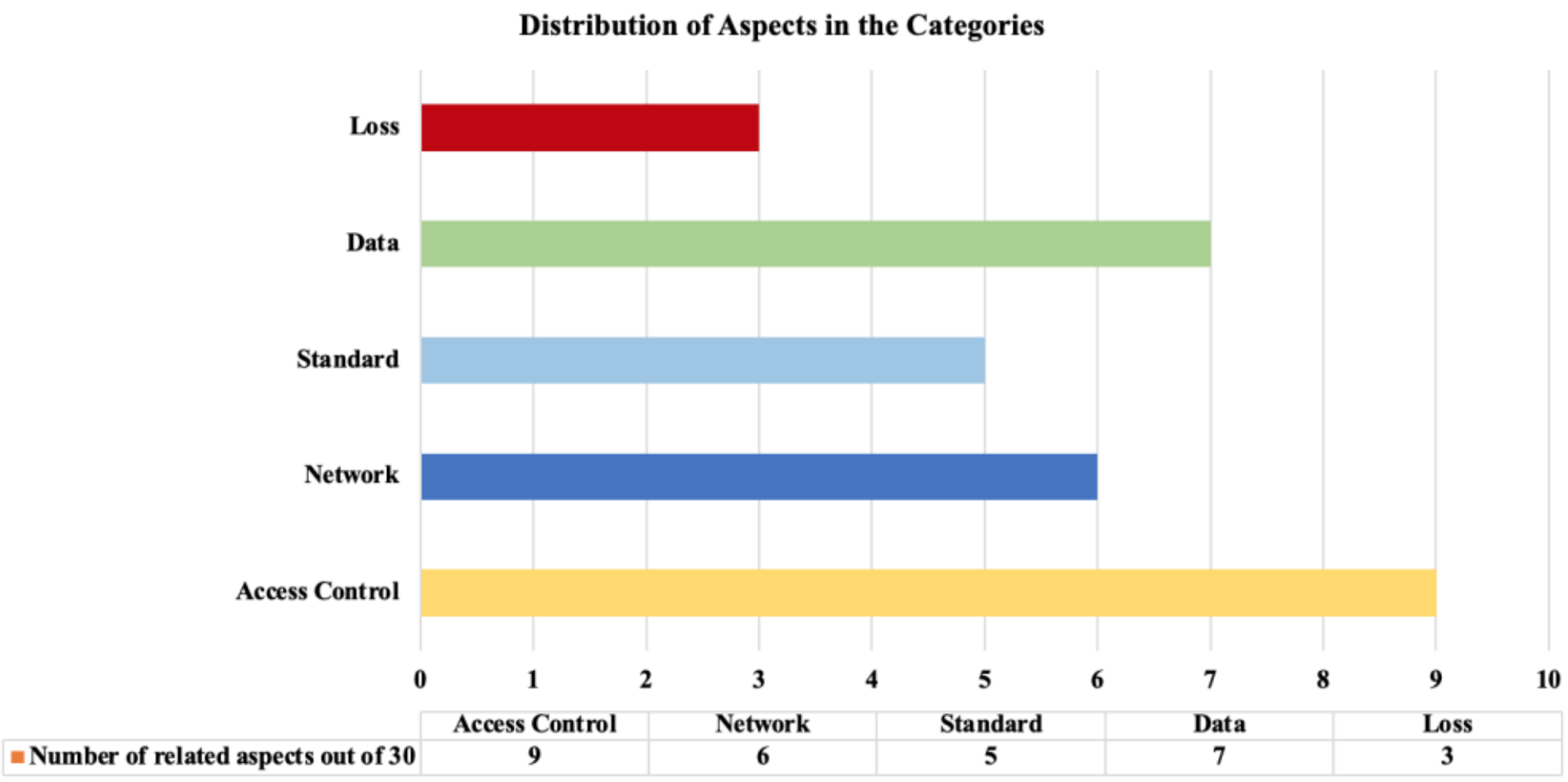

Moreover, Figure 3 shows the percentage of each of the five categories. These values are computed based on the percentage of the proportion of the number of security aspects in each category to the total number of aspects. For instance, as demonstrated in Figure 2, the access control category has nine nodes in BSAGIoT, which results 30.00% of the total security aspects. Figure 3 demonstrates that most of the security aspects are related to access control and the proper access level should be granted. Hence, we can conclude that a vast amount of the security breaches come from insufficient access control on the resources, which makes it significantly important. On the other hand, standard and network categories have a major impact on access control; therefore these categories require close attention too. Moreover, if these three categories of aspects are well managed, there will be minimum to no vulnerability and threat in the remaining categories, data and loss, that results a safe and secure IoT network. To conclude, an ideal IoT network on one hand will be free of any data breach, tampering, and/or leak, and on the other, no extortion, fraud, and/or severe and none-redeemable consequences that risks health and/or lives will exist.

**Figure 3** Percentage of security aspects based on the category type

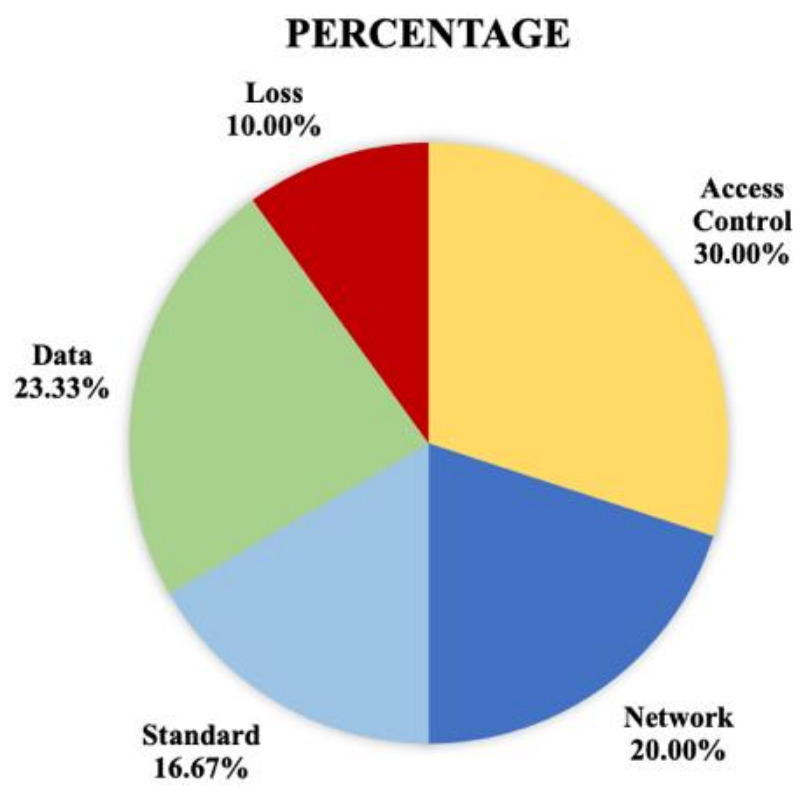

The variety of IoT applications make it a suitable target for attackers, hence preserving and maintaining its security becomes a vital issue, and BSAGIoT could be very useful for this purpose. BSAGIoT, as explained in detail in this section, could be practically used in the following conditions:

- Defining and implementing security standards and policies.
- Embedding security in IoT network's design and implementation.
- Analyzing the security of an existing IoT network either theoretically or practically.
- Performing risk assessment according to the analysis results.
- Identifying the root cause(s) and consequence(s) of security violations in an IoT environment.
- Eliminating and resolving the security issues based on the gathered information.

Aside from the mentioned attempts by organizations in designing, implementing, and using IoT networks, governments' efforts are also required to legislate and enrich the regulations. These collective efforts align with achieving the goal of secure and reliable IoT networks.

## 6 Experimental Results

Based on the definitions presented in Section 4, BSAGIoT is a BSAG that could be used for any IoT network of any application. Therefore, to alleviate or eliminate the security risks, further useful analysis could be performed according to the network's conditions. In this section, we will demonstrate how the security expert has the ability to assess the power of the potential attackers by consuming the concept of BSAG, in practice. For this purpose, the security specialist could use the introduced security aspects and their corresponding CVSS scores to estimate the conditional probability of exploiting vulnerabilities. The given guidelines in CVSS, assist the administrator to quantify the specifications of the IoT network and determine the power of the potential attackers. Furthermore, via utilizing the Bayesian inference, the security administrator could propagate the probabilities throughout the BSAGIoT and compute the unconditional probabilities of security aspects. It should also be noted that in BSAGIoT, all the nodes with multiple entry nodes or in other words multiple parents have the logical OR relation between them (equation (4)). Hence, via the usage of equations (2) and (4), the unconditional probability of the remaining nodes are also computed via propagation.

### *6.1 Case-Study*

As a case-study, we have utilized the Edge layer of the testbed presented in (Ferrag et al., 2022), a comprehensive and realistic cybersecurity dataset that accurately simulates a real-world IoT/IIoT environment and can be publicly accessed from (Edge-IIoTset dataset, 2025). The purpose is to demonstrate how these aspects impact the security and privacy of a given IoT network. Therefore, based on the network's configurations, we identified the existing vulnerabilities and obtained their CVSS metrics and Base Scores via NVD, as presented in Section 4.2. The CVSS version 3.0 user guide (CVSS, 2024) provides a comprehensive explanation of the metric values used in the allocation process. Since CVSS Base Scores range from 0 to 10 (Section 4.2), dividing them by 10 normalizes the values to a [0, 1] interval, allowing their interpretation as probabilities. As a result, the Conditional

Probability Table (CPT) entries of each aspect can be calculated using equations (2) and (4). Hence, for each aspect in BSAGIoT, details such as the respective nodes in the testbed, the related CVE-IDs, the overall eight key metrics (Section 4.2), and the CVSS Base Scores are presented in Table 4.

In Table 4, in cases where there was insufficient information, which were mainly related to the software specifications, nodes A11, A22, and A26 – A29, we utilized the expert's knowledge to quantify the score of the respective aspect. Hence, the allocated Base Scores for the mentioned aspects are 0.51, 0.6, 0.51, 0.6, 0.7, and 0.7 respectively. This indicates the security administrator's subjective belief on a successful attack.

The proposed BSAGIoT model is implemented using GeNIe Modeler, "a tool for artificial intelligence modeling and machine learning with Bayesian networks and other types of graphical probabilistic models" (GeNIe Modeler, 2024). The resulted graph, demonstrates the routes that a remote attacker could take to exploit the vulnerabilities in the network, as demonstrated in Figure 4.

### *6.2 Discussion*

With the BSAGIoT model in hand, Bayesian inference can be utilized to calculate unconditional probabilities of attackers compromising the network aspects. Hence, we present 3 scenarios to demonstrate the feasibility of the proposed method in practice, as demonstrated in Table 5. It is worth noting that the initial scenario is static and the remaining two scenarios are dynamic. This indicates that in scenario 1, the network is working with the initial conditions and probabilities where no evidence of compromise and/or validation of a node being secure has taken place. In other words, in this scenario nothing has happened and the network is running as normal. However, in the dynamic cases, during the functional time of the network evidence is found and we convey the effects of that in the model and update the probabilities dynamically. The evidence either reveals an exploited aspect or indicates one that remains unexploited, where the conditional probability of the related aspect is 1 or 0 respectively.

**Scenario 1:** In this scenario, we assume there is no specific evidence of compromise and/or validation of a node being secure has taken place; therefore, the network functions normally. In other words, in this scenario nothing has happened and the network is running as normal. As a result, via the concept of BSAG and the connections between the aspects in BSAGIoT, the unconditional probability of each node is calculated and presented in the initial column (P(Ai)) of Table 5. To conclude, in this scenario, the unconditional probabilities of compromising aspects A8, A9, and A15 as attacker's goals are 0.585, 0.585 and 0.311 respectively. This indicates a higher likelihood of financial loss, blackmail, and fraud compared to risks to health and/or life. This imbalance is somewhat expected, given the types of components in the Edge layer described by (Ferrag et al., 2022), which include soil moisture sensors, ultrasonic sensors, and pH sensors – devices primarily focused on environmental monitoring rather than human safety.

**Table 4** Existing vulnerabilities in the testbed and the related CVSS metrics and Base Scores

| ASPECT | NODE | CVE-ID | AV | AC | PR | UI | S | C | I | A | CVSS BASE SCORE |
|---|---|---|---|---|---|---|---|---|---|---|---|
| A1 | MQTT | CVE-2021-41039 | NETWORK | LOW | NONE | NONE | UNCHANGED | NONE | NONE | HIGH | 0.75 |
| A2 | RASPBERRY PI 4 Model B | CVE-2023-46837 | LOCAL | LOW | LOW | NONE | UNCHANGED | LOW | NONE | NONE | 0.33 |
| A3 | NODE-RED MODBUS TCP | CVE-2018-18759 | NETWORK | LOW | NONE | NONE | UNCHANGED | NONE | NONE | HIGH | 0.75 |
| A4 | NODE-RED MODBUS TCP | CVE-2021-3223 | NETWORK | LOW | NONE | NONE | UNCHANGED | HIGH | NONE | NONE | 0.75 |
| A5 | RASPBERRY PI 4 Model B | CVE-2017-5927 | NETWORK | LOW | NONE | NONE | UNCHANGED | HIGH | NONE | NONE | 0.75 |
| A6 | NODE-RED MODBUS TCP | CVE-2019-10756 | NETWORK | LOW | LOW | NONE | UNCHANGED | HIGH | HIGH | HIGH | 0.88 |
| A7 | NODE-RED MODBUS TCP | CVE-2019-6549 | NETWORK | LOW | HIGH | NONE | UNCHANGED | HIGH | HIGH | HIGH | 0.72 |
| A10 | RASPBERRY PI 4 Model B | CVE-2022-23960 | LOCAL | HIGH | LOW | NONE | CHANGED | HIGH | NONE | NONE | 0.56 |
| A11 | | Expert | | | | | | | | | 0.51 |
| A12 | RASPBERRY PI 4 Model B | CVE-2021-34387 | LOCAL | LOW | HIGH | NONE | UNCHANGED | HIGH | HIGH | HIGH | 0.67 |
| A13 | MQTT | CVE-2021-34431 | NETWORK | LOW | LOW | NONE | UNCHANGED | NONE | NONE | HIGH | 0.65 |
| A14 | NODE-RED MODBUS TCP | CVE-2022-3783 | NETWORK | LOW | NONE | REQUIRED | CHANGED | LOW | LOW | NONE | 0.61 |
| A16 | ESP32 | CVE-2021-34173 | NETWORK | LOW | NONE | NONE | UNCHANGED | NONE | NONE | HIGH | 0.75 |
| A17 | NODE-RED MODBUS TCP | CVE-2019-6531 | NETWORK | HIGH | NONE | NONE | UNCHANGED | HIGH | HIGH | HIGH | 0.81 |
| A18 | NODE-RED MODBUS TCP | CVE-2019-6527 | NETWORK | LOW | NONE | NONE | UNCHANGED | HIGH | HIGH | HIGH | 0.98 |
| A19 | RASPBERRY PI 4 Model B | CVE-2023-41325 | LOCAL | LOW | HIGH | NONE | UNCHANGED | HIGH | HIGH | HIGH | 0.67 |
| A20 | RASPBERRY PI 4 Model B | CVE-2021-38545 | NETWORK | HIGH | NONE | NONE | UNCHANGED | HIGH | NONE | NONE | 0.59 |
| A21 | RASPBERRY PI 4 Model B | CVE-2020-24572 | NETWORK | LOW | LOW | NONE | UNCHANGED | HIGH | HIGH | HIGH | 0.88 |
| A22 | | Expert | | | | | | | | | 0.60 |
| A23 | MQTT | CVE-2019-5432 | NETWORK | LOW | NONE | NONE | UNCHANGED | NONE | NONE | HIGH | 0.75 |
| A24 | ESP32 | CVE-2021-41104 | NETWORK | LOW | NONE | NONE | UNCHANGED | NONE | HIGH | NONE | 0.75 |
| A25 | ESP32 | CVE-2020-11015 | NETWORK | LOW | NONE | NONE | UNCHANGED | HIGH | HIGH | NONE | 0.91 |
| A26 | | Expert | | | | | | | | | 0.51 |
| A27 | | Expert | | | | | | | | | 0.60 |
| A28 | | Expert | | | | | | | | | 0.70 |
| A29 | | Expert | | | | | | | | | 0.70 |
| A30 | RASPBERRY PI 4 Model B | CVE-2021-41583 | NETWORK | LOW | LOW | NONE | UNCHANGED | HIGH | NONE | NONE | 0.65 |

**Figure 4** BSAGIoT model of the test network

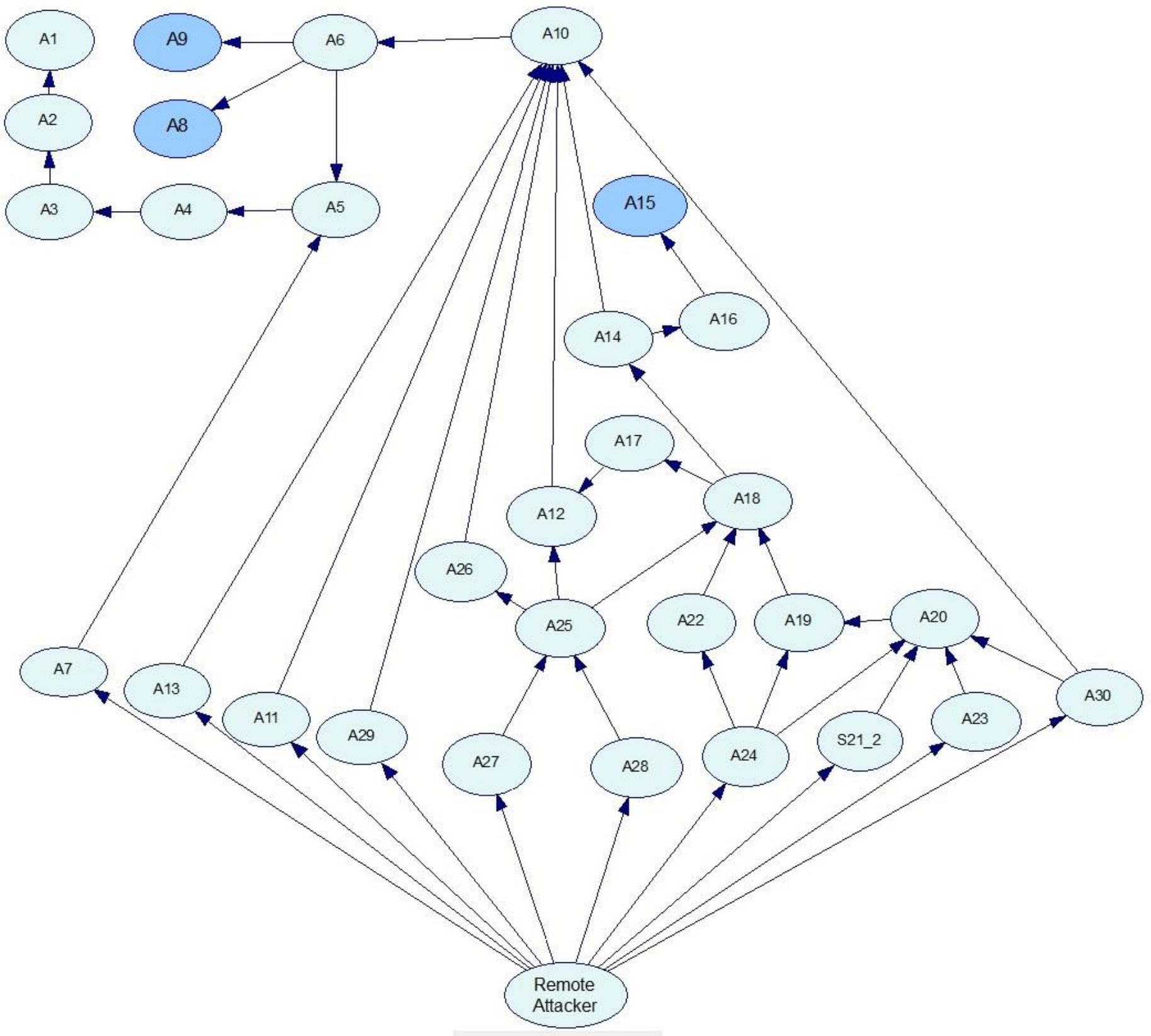

**Scenario 2:** In the second scenario, we assume evidence exists that an aspect is exploited; hence the conditional probability of the related aspect is 1. As a result, we consider a security breach is identified for aspect A25. Therefore, the probability of compromising A25 is set to 1. Accordingly, the probability of all nodes in the BSAGIoT should be updated by conducting Bayesian inference considering this evidence. As a result, the unconditional probability of each node is listed in the third column of Table 5. There is no doubt that the probabilities would either remain the same or increase in comparison to the prior scenario. Hence, the probabilities of an attacker compromising aspects A8, A9, and A15 are 0.843, 0.843, and 0.456 respectively. As expected, the probabilities of the 3 aspects aimed by the attacker are more than the original probabilities, given that we have clear evidence that A25 has been compromised, in other words P(A25) = 1.

**Scenario 3:** In the last scenario, we assume evidence exists that an aspect remains unexploited; hence the conditional probability of the related aspect is 0. As a result, we consider that aspect A23 is secure and no security breach is detected for it. Therefore, the probability of compromising A23 is set to 0. Accordingly, similar to scenario 2, the probability of all the nodes in the BSAGIoT should be updated by conducting Bayesian inference considering this evidence. As a result, the unconditional probability of each node is updated and presented in the last column of Table 5. Hence, the probabilities of an attacker compromising aspects A8, A9, and A15 are 0.308, 0.308, and 0.163 respectively. As expected, the probabilities of the 3 aspects aimed by the attacker are less than the original probabilities, given that we have clear evidence that A23 has not been compromised, in other words P(A23) = 0.

In addition, via the experiments performed in the mentioned scenarios, we conducted a feasibility study to assess the practicality of the proposed method in a real-world IoT environment. These scenarios indicate how a successful compromise and/or a failed breach could impact the overall security and privacy of the respective IoT network. In each scenario, the probabilities represent the likelihood of an aspect's security being compromised, reflecting its unconditional probability. Consequently, we can prioritize aspects based on these probabilities, with higher probabilities indicating a greater risk of compromise and a lower level of security. Therefore, aspects with the highest probabilities should be prioritized for further investigation by security experts to address the related vulnerabilities within the IoT network. Furthermore, with the results of this assessment, the security administrators and decision makers could decide which aspects should be addressed first to mitigate and/or eliminate the existing security deficiencies, which leads to risk mitigation.

**Table 5.** Unconditional Probabilities (P(Ai)) on BSAGIoT nodes in different scenarios

| NODE | P(Ai)<br>Scenario 1 | P(Ai) - A25 COMPROMISED<br>Scenario 2 | P(Ai) - A23 NOT COMPROMISED<br>Scenario 3 |
|---|---|---|---|
| **A1** | 0.081 | 0.116 | 0.042 |
| **A2** | 0.108 | 0.154 | 0.057 |
| **A3** | 0.326 | 0.467 | 0.172 |
| **A4** | 0.435 | 0.623 | 0.229 |
| **A5** | 0.580 | 0.830 | 0.305 |
| **A6** | 0.585 | 0.843 | 0.308 |
| **A7** | 0.504 | 0.720 | 0.265 |
| **A8** | 0.585 | 0.843 | 0.308 |
| **A9** | 0.585 | 0.843 | 0.308 |
| **A10** | 0.664 | 0.958 | 0.350 |
| **A11** | 0.357 | 0.510 | 0.188 |
| **A12** | 0.549 | 0.849 | 0.288 |
| **A13** | 0.455 | 0.650 | 0.239 |
| **A14** | 0.415 | 0.608 | 0.217 |
| **A15** | 0.311 | 0.456 | 0.163 |
| **A16** | 0.311 | 0.456 | 0.163 |
| **A17** | 0.551 | 0.807 | 0.289 |
| **A18** | 0.680 | 0.997 | 0.356 |
| **A19** | 0.558 | 0.797 | 0.282 |
| **A20** | 0.635 | 0.908 | 0.308 |
| **A21** | 0.616 | 0.880 | 0.324 |
| **A22** | 0.315 | 0.450 | 0.166 |
| **A23** | 0.525 | 0.750 | 0.000 |
| **A24** | 0.525 | 0.750 | 0.276 |
| **A25** | 0.585 | 1.000 | 0.308 |
| **A26** | 0.298 | 0.510 | 0.157 |
| **A27** | 0.420 | 0.695 | 0.221 |
| **A28** | 0.490 | 0.804 | 0.258 |
| **A29** | 0.490 | 0.700 | 0.258 |
| **A30** | 0.455 | 0.650 | 0.239 |

### *6.3 BSAG Example*

In this section, we demonstrate the Conditional Probability Tables (CPTs) for a selected portion of the BSAGIoT network used in our model. These CPTs are crucial for probabilistic reasoning, with each entry representing the probability of a particular node being exploited, given that its parent nodes, nodes that directly impact its security, have already been compromised. These probabilities are calculated using the CVSS Base Scores outlined in Table 4. Figure 5 illustrates the CPTs and conditional probabilities for a subgraph of nodes from Figure 1. The edges of the graph also display the unconditional probability of compromising each node, indicating the overall probability of exploiting the related security aspect. This process is known as risk assessment. For more information please refer to (Poolsappasit et al., 2012; Khosravi-Farmad & Ghaemi-Bafghi, 2020). The calculated unconditional probabilities of each security aspect enable pre-deployment prioritization for targeted security interventions. Security aspects with higher probabilities warrant more immediate attention to eliminate or reduce the associated risks. To enhance risk management precision, the potential impact of a successful vulnerability exploitation should be considered in conjunction with its probability; multiplying these values provides a more comprehensive risk score for prioritizing mitigation efforts. (Khosravi-Farmad et al., 2014).

## 7 Conclusions and Future Works

IoT promises to simplify and enhance various aspects of life. However, the exponential growth in connected devices has introduced a multitude of security challenges. Recognizing the critical importance of security, the authors have focused on this matter. We propose a classification of security aspects into five categories named data, access control, standards, network, and loss. Furthermore, the authors have formally defined the concept of BSAG and proposed a novel Bayesian dependency graph named BSAGIoT based on it. This generic model applicable to any IoT network, illustrates the impact of these security aspects and security categories on one another with the aim to maintain security. BSAGIoT is a DAG and could also be a great reference for identifying the cause of various security challenges, and their impact on IoT networks via topological sorting. Accordingly, this research could be of great use in assisting IoT security experts and/or administrators. To justify this claim, we have investigated the functionality and performance of the proposed method on a accurately simulated real-world IoT/IIoT environment in three cases. One case where the network is working with the initial conditions and probabilities and there is no evidence of a security compromise and/or validation of a node being secure, another case where there is an evidence of an attack occurring on a specific security aspect, and finally, a case where we are assured that no attack has taken place on a specific security aspect. This feasibility study demonstrates that BSAGIoT can be utilized to model the security and privacy of any IoT network, which is a crucial requirement for building trust and ensuring reliable operation.

Future research will focus on several key areas. First, we plan to develop specific security guidelines tailored to each identified IoT category. These guidelines will consider resource constraints and the heterogeneous nature of IoT networks. Second, we aim to automate the BSAGIoT model to enable secure, reliable, accurate, and trustworthy IoT deployments with minimal manual intervention. This automation relies on thoroughly developed guidelines, grounded in both theory and practice. Finally, we will explore extending the BSAGIoT model using Bayesian Decision Networks to model the effectiveness of security countermeasures in protecting IoT network assets.

**Figure 5** CPTs and probabilities of a subset of BSAGIoT

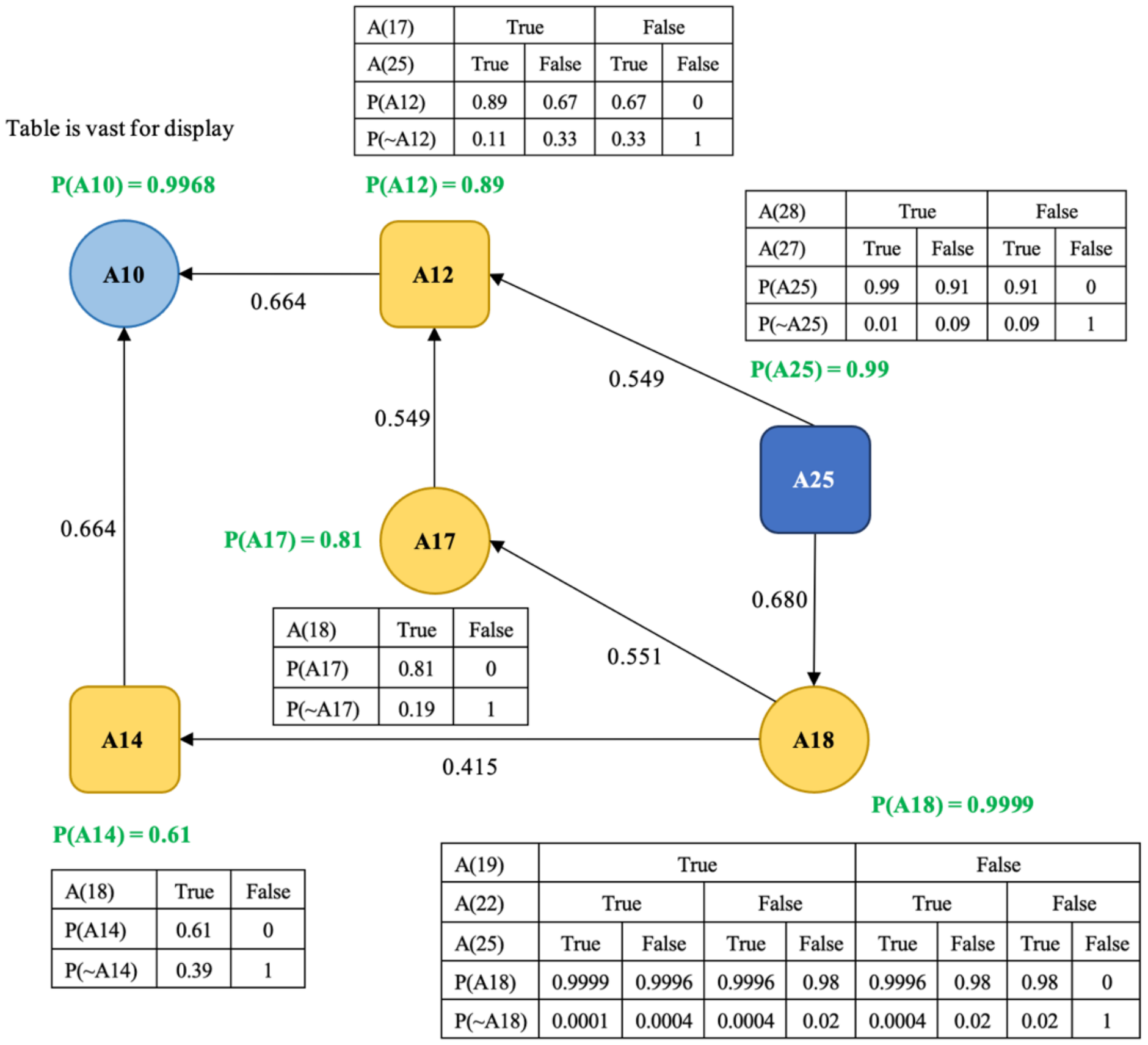

| A(17) | True | | False | |
|---|---|---|---|---|
| A(25) | True | False | True | False |
| P(A12) | 0.89 | 0.67 | 0.67 | 0 |
| P(~A12) | 0.11 | 0.33 | 0.33 | 1 |

| A(28) | True | | False | |
|---|---|---|---|---|
| A(27) | True | False | True | False |
| P(A25) | 0.99 | 0.91 | 0.91 | 0 |
| P(~A25) | 0.01 | 0.09 | 0.09 | 1 |

| A(18) | True | False |
|---|---|---|
| P(A17) | 0.81 | 0 |
| P(~A17) | 0.19 | 1 |

| A(18) | True | False |
|---|---|---|
| P(A14) | 0.61 | 0 |
| P(~A14) | 0.39 | 1 |

| A(19) | True | | | | False | | | |
|---|---|---|---|---|---|---|---|---|
| A(22) | True | | False | | True | | False | |
| A(25) | True | False | True | False | True | False | True | False |
| P(A18) | 0.9999 | 0.9996 | 0.9996 | 0.98 | 0.9996 | 0.98 | 0.98 | 0 |
| P(~A18) | 0.0001 | 0.0004 | 0.0004 | 0.02 | 0.0004 | 0.02 | 0.02 | 1 |